# Predictive modeling of stock indices closing from web search trends

Arjun R[1], Suprabha KR[2]

[1]PhD Scholar, NIT Karnataka, Mangalore- 575025
[2]Assistant Professor, NIT Karnataka, Mangalore -575025
Email: arjrs123@gmail.com, suprabha@nitk.ac.in

**Abstract.** The study aims to explore the strength of causal relationship between stock price search interest and real stock market outcomes on worldwide equity market indices. Such a phenomenon could also be mediated by investor behavior and extent of news coverage. The stock-specific internet search trends data and corresponding index close values from different countries stock exchanges are collected and analyzed. Empirical findings show global stock price search interests correlates more with developing economies with fewer effects in south asian stock exchanges apart from strong influence in western countries. Finally this study calls for development in expert decision support systems with the synthesis of using big data sources on forecasting market outcomes.

## 1 Introduction

From the past few decades, usage of big data is revolutionizing virtually on every scientific field. Big data research refers to solving specific problems by collecting and analyzing data from diverse sources including primary methods such as surveys, interviews etc. and secondary sources such as public web, machine logs, sensor data etc. In this work, we attempt showcasing some of the insightful properties using public web data for informed decision-making among stock market participants. Most of worldwide stock markets have been highly complex, stochastic and non-linear system in varying magnitudes over time (Palsson et al. 2017). Predicting or even modeling such a system with fair accuracy is highly challenging, yet one of ongoing highly researched topic in finance, economics, and computing fields (Junqué et al. 2013) (Chang and Ramachandran, 2017). The data mining tools such as Google trends and algorithm can extract predictive information about economic indicators from public internet trends (Choi and Varian, 2012). Similar domain specific data management automation tools have been proved feasible in fields of space missions (Arjun et al. 2012)

The benefits of recommendation system practically effective enough for generating profits means a lot especially for investors, markets, companies and other stakeholders who constitute this environment (Dimpfl and Jank, 2016). The additional information of public web data trends augments the forecasts of analysts even under volatility with higher accuracy and robustness over a regular expert system for managerial decision support (Attigeri et al. 2015). Hence, we assess the predictive power in search trends data to forecast the stock market index close ahead of actual transactions in India.

Two research questions are explored in this work. Research question 1: Does global stock price search affect the stock market index close at India on long-term?

Research question 2: What level of predictive power does stock price search interests have on the estimating forecasts of stock market index close?

## 2 Literature Review

Choi et al. (2012) showed how to use search engine data to forecast near-term values of economic indicators. Here simple seasonal Autoregressive integrated moving average models (ARIMA) that include relevant Google Trends variables outperform models that exclude these predictors by 5% to 20%. Preis et al. 2013 attempted in analyzing the post effect of financial markets using keywords being searched with debt as index to prove infer that information of potential market crash already existed in 2008 crisis. Dimpfl and Jank (2016) used daily search query data to measure the individual's interest in the aggregate stock market and find that investors' attention to the stock market rises in times of high market movements. Also, search queries constitute a source of information for future volatility which could essentially be used in real time. A study by Bali et al. (2017) showed that the counts of unusual news flows (either positive or negative news) accompanied by volatility shocks predicted low future returns. As quoted by Palsson et al. (2017), the dynamics of the stock market consist of a vast myriad of interacting components, whose internal details are too complex or inaccessible for one to model. Li et al. 2015 demonstrated relation between timing and approaches of macro news release and foreign exchange volatility as essential for central banks. They also stress quantifying market impact direction and magnitude of price changes arising from individual macroeconomic data releases. Also suggesting use big data sources to be used by financial regulators and market stakeholders. Liu and Yeh (2017) concluded in their study on US markets data that the evaluation of performance prediction models based on neural networks, excess return rates can be predicted more precisely than annualized return rates, total risk than systematic risk. Corea (2016) investigated three major technology companies (Apple, Facebook, Google) for two months period whether twitter platform reflects investor sentiment and suggest that negative sentiments have a negative impact on the stock price – as intuitively should be – while an increase in the posting volume of negative tweets has anyway a positive impact on the stock price. The average sentiment associated to any tweet is not relevant as expected in prediction terms, while the posting volume has a greater forecasting power. They also find that individuals with a higher influence power in social media can effectively influence the stock direction with their posts and opinions, although the magnitude is extremely low. Hence summarizing earlier works we synthesis model with two variables. Global stock price search popularity as independent variable explored by Corea 2016, Li et al. 2015, Choi et al. 2012, and Stock index close studied by Preis et al. 2013, Bali et al. 2017, Dimpfl and Jank, 2016. Also very less empirical research is found on the proposed variables within framework of Indian capital market.

## 3 Conceptual Mapping

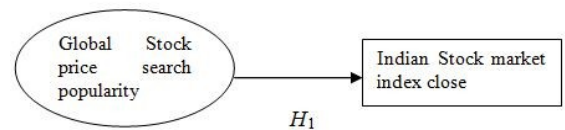

figure 1. A conceptual framework of the research study

Figure 1 shows the conceptual mapping of construct in the study.

## 4 Methodology

The study adopts triangulation approach with qualitative reasoning. The following hypothesis was formulated: Null hypothesis, $H_1$: There is no significant relation of the global stock price search interest on the Indian stock market index close.

Alternate hypothesis, $H_{1a}$: There is a significant relation of the global stock price search interest on Indian stock market index close.

## 5 Data

**5.1.** Sampling frame and Summary statistics: Stock index data- The sampling frame of the research is 2012 to 2017. Such a sample window is taken since Google trends data is available in datasets of five year slot. Moreover it partially covers financial months of corresponding years.

**Stock indices:** In current study, the highest market capitalization based country-specific indices are chosen. Such selection has been done since the search query data returns information of these specific index values. In USA- DJIA (Dow Jones Industrial Average), India- BSE 30(Bombay Stock Exchange), Singapore- STI (Strait Times Index), Hongkong- HSI (Hang Seng Index), Canada- TSX (Toronto Stock Exchange) which are all benchmark indices in respective countries. Figure 3 shows overall indices performances. Figure 3

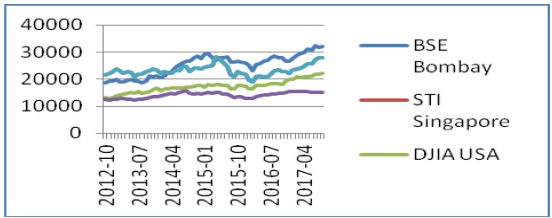

Figure 3: Performance of closing index points of chosen major stock indices during 2012-2017(Source: https://www.google.com/finance)

**Internet search patterns data:**

The sampling frame used is March 1, 2012- March 1, 2017. The web search queries and trends data are been collected from Google Trends. The user search queries on stock market index data price and location information taken. Keyword used is "S*tock price*". Figure 3 shows the Global level Stock price search popularity.

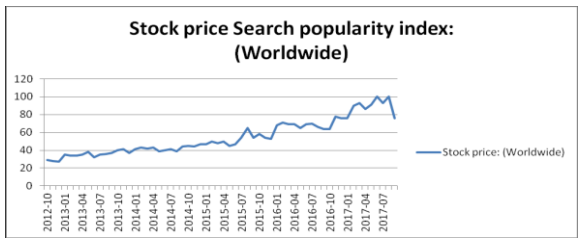

Figure 2: Interest over time (source: https://trends.google.com)

**Google Trends algorithm:** Google Trends provides an index of the volume of Google queries by geographic location and category. Google Trends data does not report the raw level of queries for a given search term. Rather, it reports a query index. The query index starts with the query share: the total query volume for search term in a given geographic region divided by the total number of queries in that region at a point in time (Choi and Varian). The query share numbers are then normalized so that they start at 0 in January 1, 2004. Numbers at later dates indicated the percentage deviation from the query share on January 1, 2004.

**Operational definition:**
**Interest over time (IOT):** Numbers represent search interest relative to the highest point on the chart for the given region and time. A value of 100 is the peak popularity of the term. A value of 50 means that the term is half as popular. Likewise, a score of 0 means term was less than 1% as popular as the peak. Algorithm to compute interest values are documented by Choi and Varian, 2012.

Table 1: Demographic distribution of countries with high Search price IOT values

| Category: All categories | |
|---|---|
| Country | Stock price: (1/1/04 - 9/13/17) |
| Singapore | 100 |
| United States | 75 |
| Canada | 68 |
| Hong Kong | 53 |
| India | 43 |
| Japan | 1 |

On levels of data, we treat the most of variables as ordinal scale. The lesser popular countries (ex. Japan etc.) are being excluded in Table 1 and from analysis.

**5.2** Analysis and methods

| Correlations | | Stock price: (Worldwide) | BSE Bombay | STI Singapore | DJIA USA | TSX Canada | HSI Hong kong |
|---|---|---|---|---|---|---|---|
| Stock price: (Worldwide) | Pearson Correlation | 1 | .764** | .246 | .888** | .576** | .246 |
| | Sig. (2-tailed) | | .000 | .058 | .000 | .000 | .058 |
| | N | 60 | 60 | 60 | 60 | 60 | 60 |
| BSE Bombay | Pearson Correlation | .764** | 1 | .543** | .900** | .828** | .543** |
| | Sig. (2-tailed) | .000 | | .000 | .000 | .000 | .000 |
| | N | 60 | 60 | 60 | 60 | 60 | 60 |
| STI Singapore | Pearson Correlation | .246 | .543** | 1 | .525** | .607** | 1.000** |
| | Sig. (2-tailed) | .058 | .000 | | .000 | .000 | .000 |
| | N | 60 | 60 | 60 | 60 | 60 | 60 |
| DJIA USA | Pearson Correlation | .888** | .900** | .525** | 1 | .804** | .525** |
| | Sig. (2-tailed) | .000 | .000 | .000 | | .000 | .000 |
| | N | 60 | 60 | 60 | 60 | 60 | 60 |
| TSX Canada | Pearson Correlation | .576** | .828** | .607** | .804** | 1 | .607** |
| | Sig. (2-tailed) | .000 | .000 | .000 | .000 | | .000 |
| | N | 60 | 60 | 60 | 60 | 60 | 60 |
| HSI Hong kong | Pearson Correlation | .246 | .543** | 1.000** | .525** | .607** | 1 |
| | Sig. (2-tailed) | .058 | .000 | .000 | .000 | .000 | |
| | N | 60 | 60 | 60 | 60 | 60 | 60 |

**. Correlation is significant at the 0.01 level (2-tailed).

**Nonparametric Correlations**

## Correlations

| Spearman's rho | | Stock price: (Worldwide) | BSE Bombay | STI Singapore | DJIA USA | TSX Canada | HSI Hong kong |
|---|---|---|---|---|---|---|---|
| Stock price: (Worldwide) | Correlation Coefficient | 1.000 | .792** | .180 | .874** | .622** | .180 |
| | Sig. (2-tailed) | . | .000 | .169 | .000 | .000 | .169 |
| | N | 60 | 60 | 60 | 60 | 60 | 60 |
| BSE Bombay | Correlation Coefficient | .792** | 1.000 | .605** | .916** | .798** | .605** |
| | Sig. (2-tailed) | .000 | . | .000 | .000 | .000 | .000 |
| | N | 60 | 60 | 60 | 60 | 60 | 60 |
| STI Singapore | Correlation Coefficient | .180 | .605** | 1.000 | .474** | .662** | 1.000** |
| | Sig. (2-tailed) | .169 | .000 | . | .000 | .000 | .000 |
| | N | 60 | 60 | 60 | 60 | 60 | 60 |
| DJIA USA | Correlation Coefficient | .874** | .916** | .474** | 1.000 | .817** | .474** |
| | Sig. (2-tailed) | .000 | .000 | .000 | . | .000 | .000 |
| | N | 60 | 60 | 60 | 60 | 60 | 60 |
| TSX Canada | Correlation Coefficient | .622** | .798** | .662** | .817** | 1.000 | .662** |
| | Sig. (2-tailed) | .000 | .000 | .000 | .000 | . | .000 |
| | N | 60 | 60 | 60 | 60 | 60 | 60 |
| HSI Hong kong | Correlation Coefficient | .180 | .605** | 1.000** | .474** | .662** | 1.000 |
| | Sig. (2-tailed) | .169 | .000 | .000 | .000 | .000 | . |
| | N | 60 | 60 | 60 | 60 | 60 | 60 |

**. Correlation is significant at the 0.01 level (2-tailed).

Analyzing the Global Stock price search interest and dependent variable as BSE index using regression analysis with a forward selection method.

**Regression**

### Variables Entered/Removed[a]

| Model | Variables Entered | Variables Removed | Method |
|---|---|---|---|
| 1 | Stock price: (Worldwide)[b] | . | Enter |

a. Dependent Variable: BSE Bombay
b. All requested variables entered.

### Model Summary[b]

| Model | R | R Square | Adjusted R Square | Std. Error of the Estimate |
|---|---|---|---|---|
| 1 | .764[a] | .583 | .576 | 2624.05017 |

a. Predictors: (Constant), Stock price: (Worldwide)
b. Dependent Variable: BSE Bombay

### ANOVA[a]

| Model | | Sum of Squares | df | Mean Square | F | Sig. |
|---|---|---|---|---|---|---|
| 1 | Regression | 558900274.2 | 1 | 558900274.2 | 81.169 | .000[b] |
| | Residual | 399367079.2 | 58 | 6885639.297 | | |
| | Total | 958267353.4 | 59 | | | |

a. Dependent Variable: BSE Bombay
b. Predictors: (Constant), Stock price: (Worldwide)

### Coefficients[a]

| Model | | Unstandardized Coefficients | | Standardized Coefficients | t | Sig. | 95.0% Confidence Interval for B | |
|---|---|---|---|---|---|---|---|---|
| | | B | Std. Error | Beta | | | Lower Bound | Upper Bound |
| 1 | (Constant) | 16675.913 | 1009.116 | | 16.525 | .000 | 14655.947 | 18695.879 |
| | Stock price: (Worldwide) | 156.419 | 17.362 | .764 | 9.009 | .000 | 121.665 | 191.172 |

a. Dependent Variable: BSE Bombay

## 6  Results and Discussions

Here, the global stock price search popularity to BSE 30 closing index showed a Pearson correlation value .764 and non-parametric based spearman correlation value (.792) which are both significant at .01 (99% confidence level) in two-tailed hypothesis tests. Also notable fact is that India is only among Asian countries taken based on search volume to show significant correlation to global trends.

The datasets and results support evidence that the global stock price search interest from online search volume is statistically somewhat related with the actual stock market index closing in countries USA, India, and Canada. The BSE stock market in India is also heavily correlated with DJIA (Dow Jones Industrial Average) US based index (.916).

Now, whether global search popularity affects Indian stock market BSE index, the bivariate regression analysis denotes the R-squared, $R^2$ = 0.583 on number of observations, N=60 and the degrees of freedom, $df$ = 58. Assuming a linear relationship, it shows the coefficient of determination of 57.6% in BSE index closing from global level stock price search popularity information. The t value 16.525 > 12.706, (t-critical value) for 2-tailed test at α=.05 significance, so alternate hypothesis $H_{1a}$ is accepted and null is rejected.

Even though there seems to be evidence of effect of global stock price search popularity in index close on long run, the predictive power for short-term forecasts gets limited as model due to higher standard error in estimation S.E= 2624.05. It's interesting to note that India is only developing economy whereas Hongkong and Singapore have no significant correlation between global stock price search trends and index closing, even though high search volume is observed during the sampling period.

**7 Conclusions**

We can conclude that based on observed increase in global stock price search interest, effects in the US, India and Canada exchanges index closing values to be higher than other countries considered. Also, STI Singapore and HSI Hongkong indices show perfect correlation $r$= 1.000, a result that requires extended research. More qualitative analysis can reveal the reasons such as investor behavior, economic factors in all these countries while development of expert systems utilizing such search trends information is important. This enhances analysts forecast estimates with a better confidence interval and reduced error estimates for the predicting market fluctuations, especially in India.

Current study intends to extend the findings with above preliminary results to reducing error residuals and capturing non-linearity with additional variables into model as shown by (Attigeri et al. 2015) and (Ho and Wang, 2016). Hence developing artificial neural network models that predict future stock index movements using such search trends information remain as further empirical research work.